# Spin selective Purcell effect in a quantum dot microcavity system

Qijun Ren,[1] Jian Lu,[1,†] H. H. Tan,[2] Shan Wu,[3] Liaoxin Sun,[1] Weihang Zhou,[1] Wei Xie,[1] Zheng Sun,[1] Yongyuan Zhu,[3] C. Jagadish,[2] S. C. Shen[1] and Zhanghai Chen,[1,*]

[1] State Key Laboratory of Surface Physics and Department of Physics, Laboratory of Advanced Materials, Fudan University, Shanghai 200433, China

[2] Department of Electronic Materials Engineering, Research School of Physics and Engineering, The Australian National University, Canberra ACT 0200, Australia

[3] National Laboratory of Solid State Microstructures, Nanjing University, Nanjing 210093, China

PACS number(s): 78.67.Hc, 78.67.Pt, 42.25.Ja, 42.50.Pq

We demonstrate the selective coupling of a single quantum dot exciton spin state with the cavity mode in a quantum dot-micropillar cavity system. By tuning an externally magnetic field, exciton spin states due to Zeeman splitting coupled differently to the cavity due to field induced energy detuning. We found a 26 times increase in the emission intensity of spin-up exciton state with respect to spin-down exciton state at resonance due to Purcell effect, which indicates the selective enhancement of light emission with the circular polarization degree up to 93%. A four-level rate equation model is developed and quantitatively agrees well with our experimental data. Our result paves the way for the realisation of future quantum light sources and the quantum information processing applications.

Novel quantum light sources and quantum information processing require the generation of light with specific linear or circular polarizations, such as polarized single photon sources [1-6] and polarization entangled photon generations [7,8]. Polarized light emissions are associated with energy states with specific spin configurations. Thus, efficient control over the energy spin states of a single quantum emitter, as well as the spontaneous emission (SE) rate, are of special interest. It has been a main prerequisite for the realization of these applications, especially with solid state implementation. A variety of methods have been proposed for controlling the SE rate of a single quantum emitter, among which, cavity quantum electrodynamic (CQED) effect, themed on the coupling between a single quantum emitter and an optical cavity, has attracted much interests [9-16]. Recently, much effort has been devoted to the single quantum dot (QD)-microcavity systems. The enhancement of SE rate of a single QD exciton state (Purcell effect) has been realized due to the efficient coupling with the microcavity, leading to controllable single photon generation [2,3]. However, manipulation of the QD-microcavity coupling in these experiments is realized using temperature or electric field tuning. These techniques do not have any control over the spin degree of freedom of excitons, leaving the underlying physics as well as the control of the SE rate of spin polarized excitons in CQED regime to remain an open question. Moreover, the traditional temperature tuning method on the cavity-QD system is slow in response, and spin dephasing is enhanced due to exciton-phonon scattering at the elevated temperature, rendering it impractical for future quantum information applications. Although quantum confined Stark effect, i.e. the electric field tuning effect, is an efficient technique for the control of the spectral resonance between a single QD and the cavity [20], complex processing procedures are required and contaminants or defects can be easily introduced into the cavity, leading to performance degradation. For the studies of the exciton spin states interacting with the cavity mode and the control of these interactions, a straightforward approach is to apply an external magnetic field. Field induced Zeeman effect lifts the degeneracy of the exciton spin states and may bring the exciton energies into resonance with the cavity mode, which would give rise to the spin selective coupling of the cavity to excitons. Although this technique has been applied in the strong coupling regime where Rabi splitting is observed [9,10], the weak coupling regime of CQED has yet to be explored. The latter regime is of great potential to demonstrate controllable circularly polarized quantum light source.

In this Letter, we present the magneto-optical investigation of the coupling between single exciton spin states of an InGaAs QD and a high quality micropillar cavity. We show that it is possible to control their coupling strength, and through spin selective Purcell effect achieve selective enhancement of the SE rate of an exciton state having a specific spin projection. The circular polarization degree of light emitted by from the QD is enhanced by up to 93% due to this effect.

The cavity QD planar microcavities were grown by metal organic chemical vapor deposition on a semi-insulating GaAs substrate. A λ-GaAs cavity is inserted between a 24-stack (top) and a 30=- stack (bottom) of highly reflective GaAs/AlAs distributed Bragg reflectors (DBRs). Self-assembled $In_{0.5}Ga_{0.5}As$ QDs were grown in the center of the cavity with the density of about $10^{10}$ cm$^{-2}$. The wafer was then processed using focused ion beam etching to form micropillar structures with circular cross section and diameters ranging from 1.5 μm to 3.5 μm. Fig. 1(a) shows the SEM image of a typical micropillar of 2.5 μm diameter. Magneto-photoluminescence (PL) studies were carried out at 37.5 K using a confocal micro-PL setup with the 632.8 nm line of a He-Ne laser as the excitation source. Due to non-resonant excitation of carriers into the GaAs matrix, the excitation power for each QD was estimated to be ~5 nW. Both excitation and collection were in the normal direction through the same objective with the laser spot size of ~ 2 μm. Magnetic field from 0 to 5 T was applied with Faraday configuration. A typical micro-PL spectrum under magnetic field of 2 T and the corresponding peak fitting are presented in Fig 1(b). Single exciton transition from one QD is observed which splits into two circularly polarized PL peaks in the presence of magnetic field due to Zeeman effect. These peaks are labeled as $X^\uparrow$ and $X^\downarrow$ which correspond to the two exciton spin states. Here, we denote the exciton state formed by an electron of spin 1/2 (-1/2) and a hole of spin -3/2 (3/2) as $X^\uparrow$ ($X^\downarrow$), respectively [10,20]. A broader peak located at the higher energy side of the spectrum is identified as the cavity mode (labeled as C). The linewidth of this peak under the excitation power of 5 nW is about 265 μeV, corresponding to a Q factor of about 4950.

Magnetic field dependent mapping of the PL spectra of these emission in Fig. 2(a) shows the crossing behavior of the of $X^\uparrow$ and C which indicates this QD-cavity system operates in the weak coupling regime. In such a case, an enhancement of the SE rate of the $X^\uparrow$ transition at zero detuning (Purcell effect) can be expected [18,20]. Both spin polarized exciton energies show quadratic behavior as a function of the magnetic field **B** due to the exciton diamagnetic shift. However, the cavity mode energy is almost unaffected, demonstrating the negligible variation of the refractive index of the GaAs and AlAs under magnetic field. Fig, 2(b) shows that by fitting the **B** dependent energies of the spin polarized exciton emissions with the function $E_\pm^X = E_0 \pm \frac{1}{2} g_{exc} \mu_B B + \gamma B^2$ (where $E_0$ is the exciton emission energy without magnetic field), we extract the exciton g factor and diamagnetic coefficient as 1.66 and 12.3 μeV·T$^{-2}$, respectively. This corresponds to total energy shift of 550 and 76 μeV for $X^\uparrow$ and $X^\downarrow$, respectively from 0 to 5 T. The different energy shifts of the excitons and cavity mode under magnetic field give rise to the crossing of the $X^\uparrow$ transition and mode resonances at around 4.0 T, while the energy detuning between $X^\downarrow$ and the cavity mode does not change significantly over the range of magnetic field investigated.

Fig. 3(a) shows the intensities of both the $X^\uparrow$ and $X^\downarrow$ transitions as a function of magnetic field. With the increase of **B**, the intensity of the $X^\uparrow$ transition also increases due to the decrease of $X^\uparrow$-C detuning ($\Delta^\uparrow$). At 4.0 T ($\Delta^\uparrow= 0$), the intensity of the $X^\uparrow$ transition reaches a peak, which is by a factor of ~ 75 higher compared to its value at low magnetic field, strongly indicating Purcell enhanced $X^\uparrow$ emission rate at the spectral resonance with the cavity mode [18,19,20]. On the other hand, since the detuning between $X^\downarrow$ and C mode ($\Delta^\downarrow$) changes very little (76 μeV) due to the competitive driving of Zeeman effect and diamagnetic shift, the intensity of $X^\downarrow$ transition is small and fairly constant over this magnetic field range. The different behaviors of the $X^\uparrow$ and $X^\downarrow$ emissions under magnetic field make it possible to control the coupling of the $X^\uparrow$ and $X^\downarrow$ states with the cavity mode by an external magnetic field. In the weak coupling regime, one can achieve selective enhancement of the emission rate for one exciton state (in our case the $X^\uparrow$ state), and hence change the relative intensities of $X^\uparrow$ and $X^\downarrow$ transitions. Note that our measurements have been done using a non-polarized and non-resonant optical excitation.

In order to quantify the QD-cavity coupling and determine the effective Purcell factor experienced by the QD, we modeled the system with four-level rate equations which include biexciton state. The use of semi-classical rate equations for describing the emission from a single QD is justified in our case, as the data were obtained from cw optical excitations. Fig. 3(b) illustrates the energy level diagram of a single QD in an external magnetic field. In our system, the QD is non-resonantly pumped with rates $P^\uparrow$ ($P^\downarrow$) for spin state $X^\uparrow$ ($X^\downarrow$), and decays by emitting photons into cavity mode with rates $F_{p,\Delta}/\tau_X$. $F_{p,\Delta}$ is defined as $F_{p,\Delta} = F_p/(1+4\Delta^2/\gamma_c^2)$, where $F_p$ is the effective Purcell factor and $\gamma_c$ is the linewidth of the cavity mode [15]. We consider the weak excitation limit, and neglect higher energy excitons, which do not manifest themselves in our experiment. The rate equations for exciton and biexciton transitions with decay lifetimes $\tau_X$ and $\tau_{XX}$, in the GaAs matrix can be described as [21,22]:

$$\frac{dP_g}{dt} = -P^{\uparrow} P_g - P^{\downarrow} P_g + \frac{\gamma + F_{p,\Delta^{\uparrow}}}{\tau_X} P_{X^{\uparrow}} + \frac{\gamma + F_{p,\Delta^{\downarrow}}}{\tau_X} P_{X^{\downarrow}},$$

$$\frac{dP_{X^{\uparrow}}}{dt} = P^{\uparrow} P_g - \frac{\gamma + F_{p,\Delta^{\uparrow}}}{\tau_X} P_{X^{\uparrow}} - P^{\downarrow} P_{X^{\downarrow}} + \frac{\gamma}{\tau_{XX}} P_{XX} - \frac{1}{\tau_{flip}} P_{X^{\uparrow}} + \frac{1}{\tau_{flop}} P_{X^{\downarrow}},$$

$$\frac{dP_{X^{\downarrow}}}{dt} = P^{\downarrow} P_g - \frac{\gamma + F_{p,\Delta^{\downarrow}}}{\tau_X} P_{X^{\downarrow}} - P^{\downarrow} P_{X^{\downarrow}} + \frac{\gamma}{\tau_{XX}} P_{XX} + \frac{1}{\tau_{flip}} P_{X^{\uparrow}} - \frac{1}{\tau_{flop}} P_{X^{\downarrow}},$$

$$\frac{dP_{XX}}{dt} = -2\frac{\gamma}{\tau_{XX}} P_{XX} + 2P^{\downarrow} P_{X^{\downarrow}},$$

where $P_g$ is the occupation probability for the exciton ground state (empty dot), $P_{X^{\uparrow}}$ and $P_{X^{\downarrow}}$ are the probabilities of having a single exciton with -1 spin (spin-up) and +1 spin (spin-down) configurations, respectively, $\tau_{flip}$ and $\tau_{flop}$ are the spin flip-flop transition time. We assume the PL measurement at saturation has pumping rates of $P^{\uparrow} \propto \sqrt{\gamma + F_{p,\Delta^{\uparrow}}}$ and $P^{\downarrow} \propto \sqrt{\gamma + F_{p,\Delta^{\downarrow}}}$ [19, 20]. Here we note that in our case the measured cavity mode linewidth $\gamma_c$ is actually larger than the value of a bare cavity by a factor of $\sqrt{\gamma + F_p/\gamma}$, due to the low pumping power [28]. We take the experimentally measured cavity linewidth $\gamma_c = 265~\mu eV$, estimated values of $\gamma \approx 1$, $\tau_X = 2\tau_{X_2} \approx 1$ ns into account [20,25,26], and calculate the emission intensities $I_{X^{\uparrow}}$ ($I_{X^{\downarrow}}$), which are proportional to $F_{p,\Delta^{\uparrow}}/(\sqrt{\gamma + F_{p,\Delta^{\uparrow}}} + \tau_x/\tau_{flop})$ ($F_{p,\Delta^{\downarrow}}/(\sqrt{\gamma + F_{p,\Delta^{\downarrow}}} + \tau_x/\tau_{flip})$) for each applied magnetic field, by solving these rate equations. The calculated results are shown in Fig. 3(a) as solid lines, which fit well with our experimental result.

More precisely, by studying the relative intensity $I_{X^{\uparrow}}/I_{X^{\downarrow}}$, the experimentally induced error bar can be minimized. We define the circular polarization degree as $P_c = (I_{X^{\uparrow}} - I_{X^{\downarrow}})/(I_{X^{\uparrow}} + I_{X^{\downarrow}})$. The polarization degree $P_c$ as a function of the magnetic field is presented in Fig. 4(a). Remarkably, it demonstrates a well pronounced peak corresponding to the resonance between the upper spin polarized exciton state and the cavity mode. The decrease of the circular polarization degree with further increase of the magnetic field is a direct manifestation of Purcell effect. The results of the model are in a quantitative agreement with the experimental results. The best fitting yields an effective Purcell factor of 5.9±2.0, and a circular polarization degree up to 93% at a magnetic field of 4.0 T. We found the time constant for the spin flip-flop process is about tens of times longer than $\tau_X$, and this result is consistent with the existing measurement[30]. Across the magnetic field range studied here, their intensity ratio $I_{X^\uparrow}/I_{X^\downarrow}$ as well as the circular polarization degree can be continuously tuned, enabling the control on polarization selectivity of a single QD spin state emission.

To further explore the properties of coupling between the exciton spin states and the cavity, the PL linewidths of the $X^\uparrow$ and $X^\downarrow$ emissions with respect to magnetic field are plotted in Fig. 4(b). It is obvious the magnetic field dependence of the linewidths is similar to that of the intensities plots. For the $X^\uparrow$ transition, the linewidth increases from 80 μeV (with $\Delta^\uparrow$= -388 μeV at 0 T) to about 250 μeV at 4.0 T (with $\Delta^\uparrow$= 0) and then decreases to 105 μeV at a magnetic field of about 5.0 T. For non-resonant excitation, the homogeneous broadening of single QD exciton transition can be expressed as $T_2^{-1} = (2T_1)^{-1} + (T_2^*)^{-1}$, where $T_1$ is the exciton radiative lifetime and $T_2^*$ is the pure dephasing time without recombination [23]. The exciton radiative lifetime is modulated by the coupling with cavity mode due to Purcell effect. In addition the radiative lifetime can be expressed as a Lorentzian function of the detuning $\Delta^\uparrow$ between the cavity mode and the emitter. Therefore, we attribute the variation of the emission linewidth to the modification of the radiative lifetimes of spin polarized excitons by Purcell effect in the weak coupling regime. With the increase of magnetic field, the decay rate of $X^\uparrow$ emission is enhanced and the corresponding linewidth, which is inversely proportional to the decay rate, is broadened due to the decrease of $\Delta^\uparrow$. After the $X^\uparrow$ energy crosses the cavity mode, the linewidth is narrowed again, showing the revival of its decay rate at large detuning. For $X^\downarrow$ emission, the evolution of its linewidth is not as obvious as that of $X^\uparrow$ due to its small energy change with applied magnetic field . We notice the linewidth of $X^{\downarrow}$ also increases slightly from 80 μeV to 89 μeV as the $X^{\uparrow}$ energy crosses with the cavity mode, then decreases to 82 μeV at 5T. The comparison of linewidths of both of the $X^{\uparrow}$ and $X^{\downarrow}$ emissions strongly demonstrates the coupling between the two exciton spin states and the cavity mode, and shows how the emission intensities of these states can be manipulated by an external magnetic field.

In summary, we observe the coupling of single QD exciton spin states with a micropillar cavity controlled by magneto-optical tuning of their spectral energies. Purcell effect for both of spin polarized exciton emissions was observed and analyzed with a four level scheme. The experimental results as well as the theoretical simulations demonstrate that excitons with specific spin configuration can be selectively coupled to light and their decay rates can be controlled by magnetic manipulation. Our results demonstrate the feasibility of controlling the polarization of single QD emission. We achieve photon emission with circular polarization degree up to 93%, which paves the way for the realization of future quantum light sources and quantum information applications.

The work is funded by the NSF and 973 projects of China (No.2011CB925600). The authors are grateful to the ARC and ANFF for sample growth and fabrication and Professor Alexey Kavokin for stimulating discussion.

**Figures Caption:**

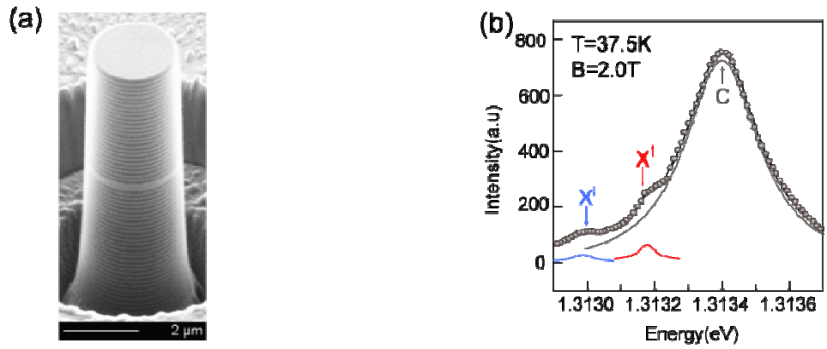

Fig. 1 (a) Scanning electron microscopy (SEM) image of a micropillar cavity with a diameter of 2.5 μm. (b) PL spectrum of the pillar with the same size as shown in the SEM image under a magnetic field of 2 T. The splitting of single exciton transition of a single QD to $X^{\uparrow}$ and $X^{\downarrow}$, as well as the fundamental mode of the micropillar cavity, C, are identified. Solid lines show the result of a multi-Lorentzian fitting.

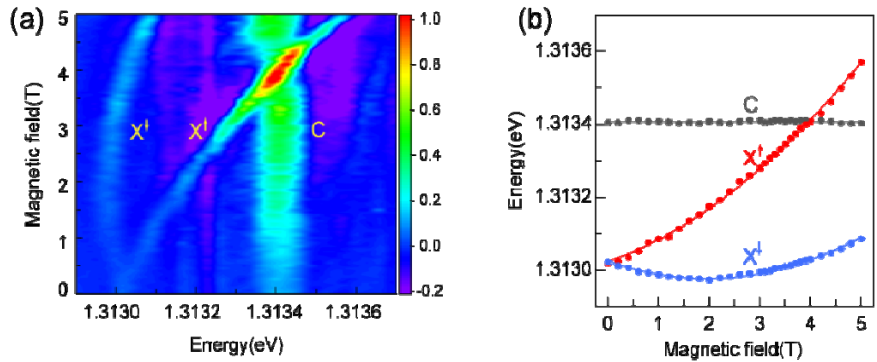

Fig. 2(a) Magnetic field map of the PL spectra from $X^{\uparrow}$, $X^{\downarrow}$ and C emissions. (b) Detunings of $X^{\uparrow}$ and $X^{\downarrow}$ emissions with respect to the cavity mode as a function of applied magnetic field. Solid lines are fitting curves, where the exciton g factor and diamagnetic parameter can be extracted.

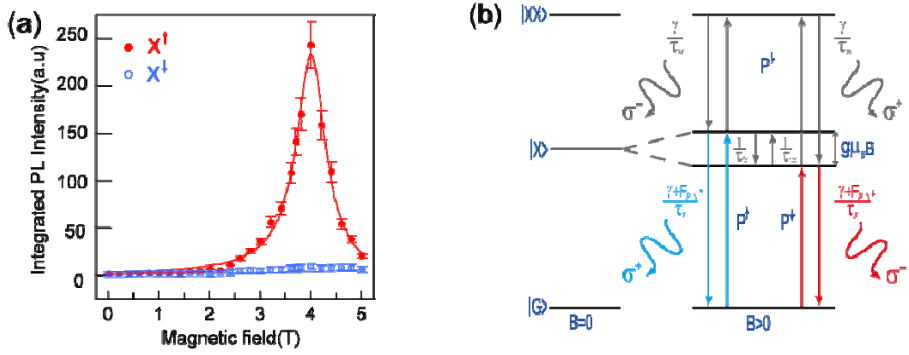

Fig.3 (a) Integrated intensities of $X^\uparrow$ (red) and $X^\downarrow$ (blue) emissions obtained from Lorentzian fittings of the spectra under different magnetic fields. Corresponding solid lines show the calculation results based on four-level rate equations. (b) Diagram of the transition between single exciton states under magnetic field, from which the four-level rate equations are developed.

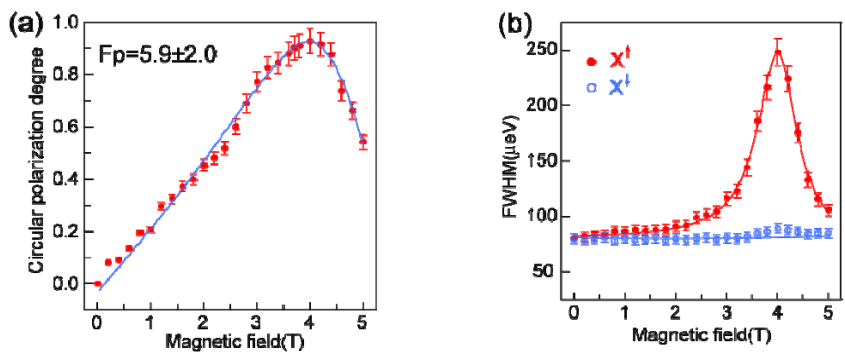

Fig. 4 (a) Circular polarization degree as a function on magnetic field. The solid line is the theoretical calculations based on four-level rate equations. (b) Linewidths of $X^\uparrow$ (red) and $X^\downarrow$ (blue) emissions as a function of magnetic field. The solid lines are just a guide to the eye.